\documentclass[12pt,a4paper]{article}
\usepackage[utf8]{inputenc}
\usepackage{t1enc}
\usepackage{hyperref}
\usepackage{times}
\usepackage{graphicx}
\usepackage{amsmath}
\usepackage{stmaryrd}
\usepackage{cite}
\usepackage{xcolor}
\usepackage{geometry}
\title{Why integral equations should be used instead of differential equations to describe the dynamics of epidemics}

\author
{Zoltan Fodor,$^{1,2,3,4}$ Sandor D. Katz,$^{2}$ Tamas G. Kovacs$^{2,5}$\\
\\
\normalsize{$^{1}$Department of Physics, University of Wuppertal, D-42119 Wuppertal, Germany}\\
\normalsize{$^{2}$ELTE Eotvos Lorand University, H-1117 Budapest, Hungary}\\
\normalsize{$^{3}$Julich Supercomputing Centre, Forschungszentrum Julich, D-52428 Julich, Germany}\\
\normalsize{$^{4}$University of California, San Diego, La Jolla, CA 92093, USA}\\
\normalsize{$^{5}$Institute for Nuclear Research, H-4026 Debrecen, Hungary}\\
}

\date{}

\begin{document}

\maketitle

\begin{abstract}
It is of vital importance to understand and track the dynamics of rapidly
unfolding epidemics. The health and economic consequences of the current
COVID-19 pandemic provide a poignant case. Here we point out that since they
are based on differential equations, the most widely used models of epidemic
spread are plagued by an approximation that is not justified in the case of
the current COVID-19 pandemic. Taking the example of data from New York City,
we show that currently used models significantly underestimate the initial
basic reproduction number ($R_0$). The correct description, based on integral
equations, can be implemented in most of the reported models and it much more
accurately accounts for the dynamics of the epidemic after sharp changes in
$R_0$ due to restrictive public congregation measures. It also provides a
novel way to determine the incubation period, and most importantly, as we
demonstrate for several countries, this method allows an accurate monitoring of
$R_0$ and thus a fine-tuning of any restrictive measures. Integral equation based
models do not only provide the conceptually correct description, they also have 
more predictive power than differential equation based models, therefore we do not
see any reason for using the latter. 
\end{abstract}

\section*{Introduction}

A commonly used approach to describe the dynamics of epidemics is based on SEIR-type 
(Susceptible-Exposed-Infectious-Recovered) differential equations~\cite{anderson1992infectious,
lipsitch2003transmission,wu2011use,fraser2009pandemic,klepac2009stage}. Recently these methods
have been applied to the COVID-19 pandemic to determine the basic reproduction number~\cite{wu2020nowcasting,tang2020estimation,tang2020updated,li2020early,wang2020phase,kucharski2020early}, the incubation period~\cite{lauer2020incubation,backer2020incubation,
linton2020incubation} and to describe the dynamics of the pandemic~\cite{chinazzi2020effect,
kraemer2020effect,prem2020effect,li2020substantial,yang2020modified,
boldog2020risk,anderson2020will,Kisslereabb5793}.
In this framework there is no natural and transparent 
way to account for the delay between the introduction of public health measures and the corresponding change in the newly reported cases. This delay occurs due to the incubation period of 
the disease. Even after refinements to try to account for some aspects 
of this delay~\cite{Wang2018Characterizing,chinazzi2020effect,kraemer2020effect,prem2020effect}, these models are still plagued by uncontrolled 
approximations. We suggest that models based on differential equations 
should be replaced with ones based on an 
integral equation, which explains the time delay without any additional step or input. 
The approach we present is not completely new, it was already implicit 
in the original Kermack-McKendric theory proposed in 1927
\cite{kermack1927contribution}. Even though several variants of that model with various 
degrees of precision circulate in the current literature
\cite{hethcote1980integral,wearing2005appropriate,wallinga2007generation,
flaxman2020report}, their superiority over the differential equation based method 
has not been widely recognised.

We emphasise that in the form presented below, the integral equation based 
approach is neither technically nor conceptually more complicated than 
the one based on differential equations. More importantly, it gives a much 
more realistic representation of the epidemic dynamics. We demonstrate 
this by comparing how the two types of approach describe the New York 
City data~\cite{nyc_data}, taken from the currently ongoing COVID-19 pandemic. Firstly, 
we show that even in the initial simple exponential phase of the 
epidemic, the correct, integral equation based model can give a 
significantly 
larger estimate for $R_0$, the basic reproduction number. Secondly, we 
show that the oscillations in the graph of the number of newly infected 
people after restrictive public interaction measures are introduced, is well described by 
the integral equation model. This is a generic feature of COVID-19 data, 
seen in several countries, and there is no simple way the differential 
equation based models could explain it. Thirdly, using data of several countries, we 
illustrate how the change of $R_0$ in time can be monitored using the integral equation formalism.

Our comparison here is based on a simple SEIR-type differential equation 
model and its integral equation counterpart containing the same 
variables, but a different time evolution. It can be seen from our 
general discussion that the differences we find are generic features of 
the two types of models and carry over to more sophisticated versions thereof.
We urge everyone to test how different their results 
are between the two approaches in the case of the particular models they use.

\section*{Integral equation description for discrete and continuous time evolution}

\begin{figure}[t]
\begin{center}
\includegraphics[width=0.5\textwidth]{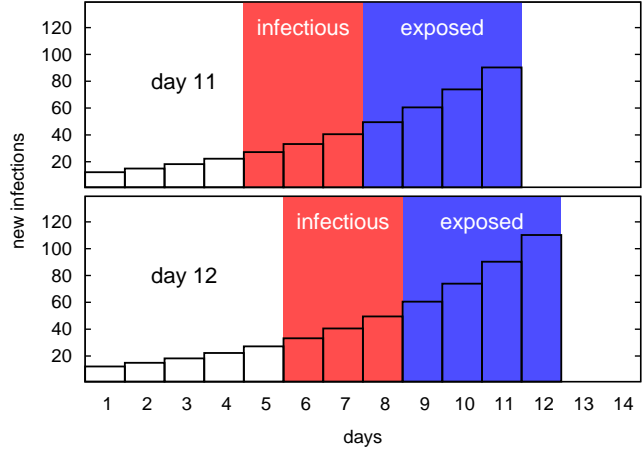}
\caption{The number of new infections per day, shown up to the eleventh (top)
  and the twelfth day (bottom) of the epidemic. The blue area represents those
  that are infected, but not yet infectious, i.e.\ people who were exposed
  between today and $t_e=4$ days ago. The red area represents those that are
  infectious, i.e.\ people who were exposed more than $t_e$ days ago, but not
  more than $t_e+t_i=7$ days ago. In the SEIR model the number of people who
  cease to be infectious from day 11 to day 12 is equal to the daily average
  of the red area in the top panel. In reality, however, only people in the
  leftmost red histogram, here a much smaller number, cease to be
  infectious. Similarly, in the SEIR model the number of people who become
  infectious is the daily average of the blue area, when in reality it is only
  people in the leftmost bar of the blue area who become infectious from day
  11 to day 12. Clearly the only case when the SEIR model correctly
  represents the situation is when the number of newly infected people per day
  is constant in time.
\label{fig:hist}
}
\end{center}
\end{figure}

The mathematical modelling of how infectious diseases spread is almost
exclusively based on compartmental models. In this framework the population is
divided into different categories and a dynamical model is set up to describe
how the number of individuals in each category evolves with time. A simple
model of this type is the so called SEIR model, in which the compartments are
susceptible (not yet infected), exposed and infected (but not yet capable of
infecting others), infectious, and recovered/removed (not capable of infecting others
any more). In its simplest form the model is characterised by three parameters,
$\alpha, \beta, \gamma$ that determine the transition rates among different
compartments through the following set of differential equations:
\begin{equation}
  \frac{dS}{dt}  =  -\beta \frac{I}{N} S, \;\;\;\;\;
  \frac{dE}{dt}  =  \beta \frac{I}{N} S - \alpha E, \;\;\;\;\;
  \frac{dI}{dt}  =  \alpha E - \gamma I, \;\;\;\;\;
  \frac{dR}{dt}  =  \gamma I,
\end{equation}  
where $S(t),E(t),I(t)$ and $R(t)$ are the number of susceptible, exposed,
infectious and recovered individuals, all functions of the time, and
$N=S+E+I+R$ is the total population.

In the initial phase of a rapidly developing epidemic, the situation we are
concerned with here, $S/N \approx 1$, which we will assume.  With this approximation the two functions
describing the dynamics are $E(t)$ and $I(t)$. At this point it is instructive
to introduce $t_e$ and $t_i$, the average days an individual spends in the
categories $E$ and $I$ respectively. It is convenient to write the equations
in terms of these parameters and the basic reproduction number, $R_0$, using the
simple relations $\alpha = 1/t_e$, $\beta = R_0/t_i$ and $\gamma
= 1/t_i$.

With these new parameters the simplified form of the differential equations
valid for the initial stage when $S/N \approx 1$ is
\begin{equation}
\frac{d}{dt}
\begin{pmatrix} E\\ I \end{pmatrix} =
\begin{pmatrix}
-1/t_e & R_0/t_i\\ 1/t_e & -1/t_i \end{pmatrix}
\begin{pmatrix} E \\ I \end{pmatrix}.
\label{eq:seir}\end{equation}
The SEIR equations are based on the assumption that the transition rates between compartments
are proportional to the number of people in those compartments. Furthermore, all people in a compartment
are treated equally, irrespective of how much time they already spent there.
These assumptions are only correct when $R_0$ is close to 1 and changes slowly. In the following we derive the integral equation description of epidemics.

Let us denote the number of newly infected individuals on day $t$ by
$\rho(t)$. The function $\rho(t)$ contains the entire history of infections. This can be depicted
in a histogram with time flowing from left to right along the horizontal axis
(Fig.\ \ref{fig:hist}). Based on this information we would like to determine
$\rho(t+1)$, the number of individuals becoming infected on the following
day. As in the SEIR model above let us here also assume that anyone exposed to the
infection will become infected, but will not be infectious for $t_e$
days. Such a person will become infectious $t_e$ days after exposure and will
remain in this category for an additional $t_i$ days, after which he is
isolated and ceases to be capable of infecting others. In the simplest version
of the model $t_e$ and $t_i$ can represent averages, but later on we will
indicate how to generalise the model by using continuous distributions. In
Fig.\ \ref{fig:hist} the blue area represents those in category $E$, with
$t_e=4$, and the red area those in $I$, with $t_i=3$. Now, exactly as in the
SEIR model above, the number of newly infected on the following day,
day 12 in the figure, will be
\begin{equation}
  \rho(t+1) = \beta  \sum_{\tau=t-t_i-t_e+1}^{t-t_e} \rho(\tau),
  \hspace{6ex} \beta = \frac{R_0}{t_i}
     \label{eq:discretized}
\end{equation}
where the sum is the total number of infectious individuals at time $t$,
i.e.\ the red area in the top panel of Fig.\ \ref{fig:hist}.  At this point we
can depict the situation at time $t+1$ simply by shifting both the blue and
the red region, showing the people in category $E$ and $I$, one day forward to
the right (bottom panel of Fig.\ \ref{fig:hist}). From this point
on the same procedure can be further iterated day by day: on each day we
calculate the number of newly infected for the following day and shift the
windows delineating the exposed and the infectious by one day forward.

At this point the objection could be raised that we have been comparing the
continuous SEIR model with a discretised model. However, the framework
presented in the histograms can be easily generalised from days to arbitrarily
fine time steps. In the limit of infinitely fine time steps $dt$, we can choose
the height of the histogram bars $\rho(t)$ such that $\rho(t) dt$ be the
number of people becoming infected in the time interval $[t,t+dt]$. In this
case the sum in the right hand side of Eq.\ (\ref{eq:discretized}) becomes an
integral and the equation can be rewritten as
\begin{equation}
  \rho(t) = \beta \int_{t-t_i-t_e}^{t-t_e} \rho(\tau) \, d\tau,
  \hspace{6ex} \beta = \frac{R_0}{t_i}.
   \label{eq:integ}
\end{equation}
We emphasise that only this continuous version of the model describes the real
situation properly. 

\begin{figure}[th]
\begin{center}
\includegraphics[width=0.5\textwidth]{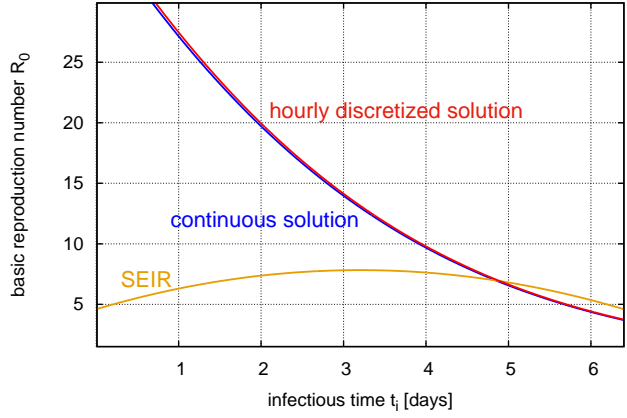}
\caption{$R_0$ as a function of
$t_i$, the infectious part of the incubation period. The total 
incubation period is taken to be $t_e+t_i$=6.4 days~\cite{backer2020incubation} and the exponential growth parameter is
$1/\lambda=1.78$ days which describe the growing phase of the New York City data~\cite{nyc_data} well.
The orange colour curve with the maximum shows the prediction (\ref{eq:analytic}) based on 
the SEIR model of eq.~(\ref{eq:seir}). The blue and red lines correspond
to the analytic solution (\ref{eq:analytic}) of the integral equation (\ref{eq:integ}) and its discretised version
(\ref{eq:discretized}) with an hourly time step, respectively. Their agreement indicate that such a discretisation is sufficient. A daily discretisation would give 34\% higher $R_0$ values.
\label{fig:conti}
}
\end{center}
\end{figure}

\section*{Determination of $R_0$, the example of New York City}

We illustrate the shortcomings of the SEIR model in two examples. First let us consider the case
of constant $R_0>1$ which leads to the exponentially growing solution with exponent $\lambda$, 
i.e. all relevant quantities ($E(t),I(t)$ of SEIR and $\rho(t)$ of the integral equation) are proportional to 
$\exp(\lambda t)$. A simple substitution gives:
\begin{equation}
R_0^{(\mbox{\scriptsize SEIR})}=1+\lambda(t_e+t_i)+\lambda^2\cdot t_e t_i, \quad
R_0^{(\mbox{\scriptsize integral})}=\frac{t_i\lambda}{e^{-\lambda t_e}-e^{-\lambda 
(t_e+t_i)}}, \quad
R_0^{(\mbox{\scriptsize discretised})}=\frac{e^{\lambda\Delta t}-1}{\lambda\Delta t} R_0^{(\mbox{\scriptsize integral})},
\label{eq:analytic}
\end{equation}
where $R_0^{(\mbox{\scriptsize discretised})}$ is the solution of the discretised equation (\ref{eq:discretized}) with time step $\Delta t$. Figure \ref{fig:conti} shows these three $R_0$ (SEIR, integral equation, discretised integral equation) obtained using realistic parameters that describe the initial phase of the pandemic in New York City. There is a striking difference between the SEIR and integral equation predictions. While the SEIR result is largely insensitive to the split of the total incubation period into $t_e$ and $t_i$, this is not the case for the solution of the 
integral equation. The true value of $R_0$ can be 3-4 times larger than the one predicted by the SEIR model. For $t_i\approx 2$ days which, as we will later see, provides a reasonable description of the daily reported cases, $R_0$ can be as high as 20. $R_0$ is, of course region dependent. Less populated areas can have much smaller $R_0$ values.

One might think that this difference is only due to the relatively large incubation period (as compared to the characteristic growth time of the pandemic) and the SEIR model gives reliable estimates for small incubation periods. Surprisingly this is not the case. A trivial expansion of $R_0^{(\mbox{\scriptsize integral})}$ in $\lambda t_e$ and $\lambda t_i$ yields $R_0^{(\mbox{\scriptsize integral})}\approx 1+\lambda(t_e+t_i/2)
+ \lambda^2(t_e^2/2+t_et_i/2+t_i^2/12)+{\cal O}(\lambda^3)$
which is clearly different from the SEIR result.

\section*{The effect of decreasing $R_0$, delay and oscillation}

\begin{figure}[th]
\begin{center}
\includegraphics[width=0.4\textwidth]{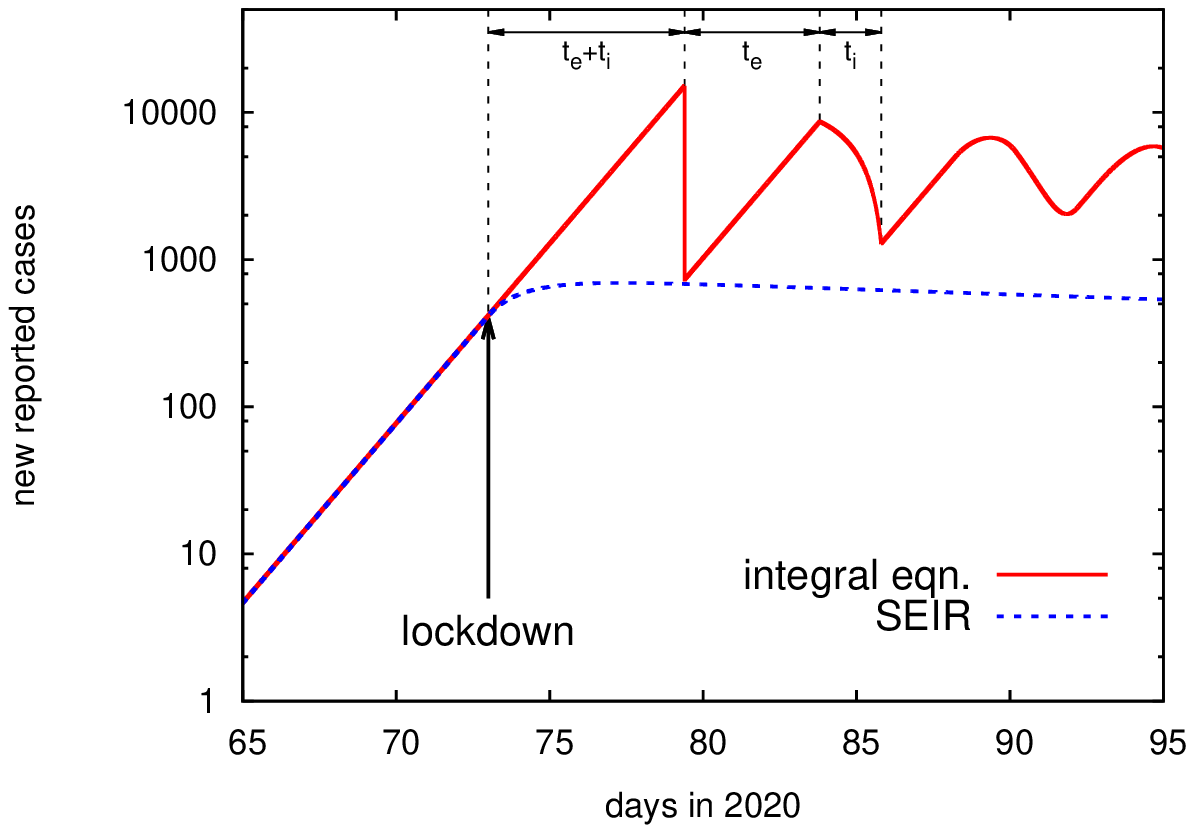}
\includegraphics[width=0.4\textwidth]{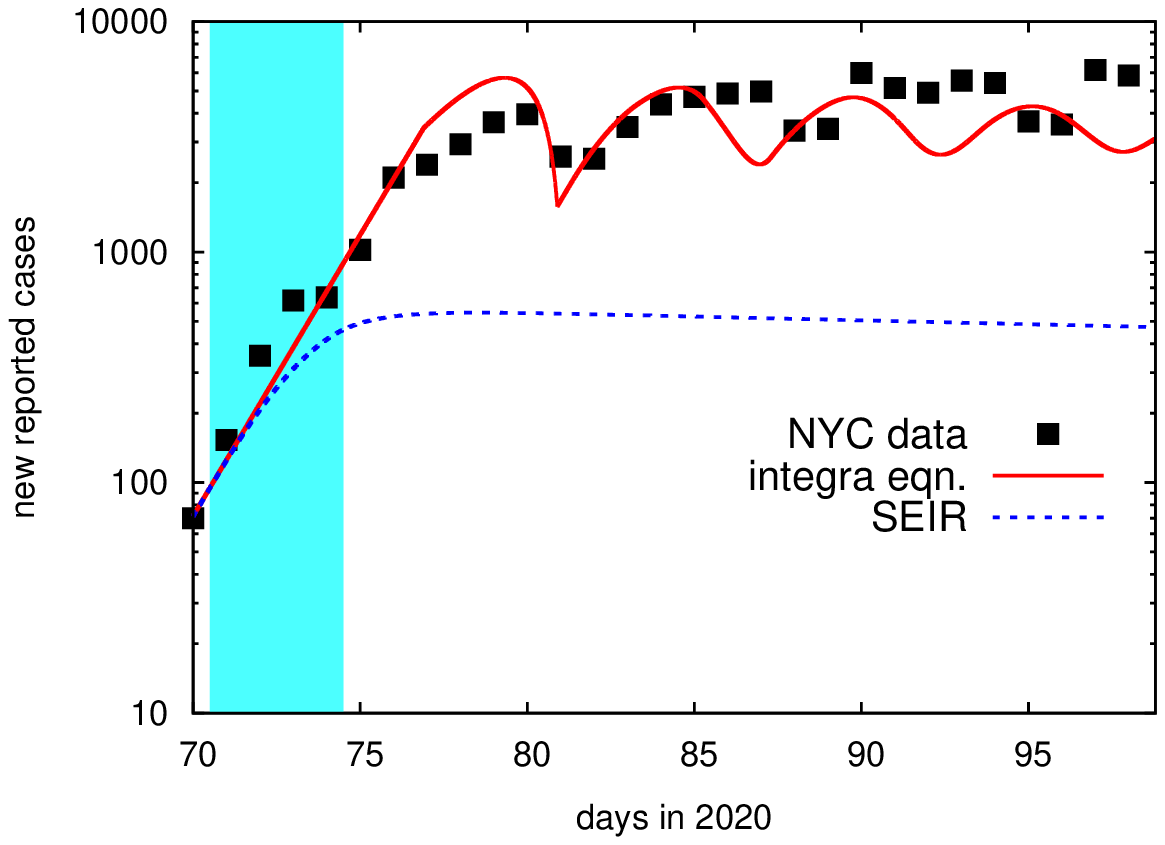}
\caption{{\it Left:} time evolution of new reported cases after a lockdown on day 73 in the case of the SEIR and integral equation models. For the initial phase of the pandemic $1/\lambda=1.78$ days, $t_e=4.4$ days and $t_i$=2 days was used. This corresponds to $R_0^{\mbox{\scriptsize SEIR}}=7.37$ and $R_0^{\mbox{\scriptsize integral}}=19.7$. The value of $R_0$ was instantaneously decreased to $0.95$ in both cases on day 73. The difference between the two models is quite dramatic. The SEIR solution reacts immediately and turns smoothly to a decreasing exponential function. The solution of the integral equation is qualitatively different. There is a delay of $t_e+t_i$ after which the effect of the lockdown becomes visible, then an oscillation follows. The characteristic scales of this oscillation are indicated in the figure. {\it Right:} time evolution of the new reported cases after a gradual lockdown from day 71 to day 74 confronted with the data of New York City. The initial parameters are the same as for the left panel. $R_0$ is now decreased to $0.95$ linearly in time during this four day period in both models (for a detailed analysis how $R_0$ changed and dropped to a smaller value in New York City see later). The main features of the curves are similar to the left panel but the amplitude of the oscillation is reduced, making it similar to the real data. Note that we did not fit our model to the data. This plot is an illustration that the main features (delay and oscillation) of the data are well captured by the integral equation and neither of these is reproduced by SEIR.}
\label{fig:evolution}
\end{center}
\end{figure}

The parameter $R_0$ in both the SEIR- and the integral equations is in general time dependent. The most important
goal of the first restrictive measures was to decrease the value of $R_0$ as fast as possible and bring it below the critical value of $1$. In the following we study how the different models react to a sudden drop of $R_0$ and
compare them qualitatively to the daily reported new cases in New York City~\cite{nyc_data}. To simplify our discussion we will assume that all cases are reported exactly after the incubation period, i.e. $t_e+t_i$ days after first exposure.
In the case of the SEIR model the rate of people leaving the incubation period is $dR/dt=\gamma I$ while in the
case of the integral equation it is $\rho(t-t_e-t_i)$. These two functions will be referred to as ''new reported cases'' in the following. If we assume that the majority of cases are reported when symptoms emerge, it follows that any change in the parameters will only show up $t_e+t_i$ days later in the data. Any reasonable model should be able to naturally account for this delay. Quantities in differential equations react immediately to any change of the parameters, thus no differential equation based model (such as SEIR or its simple extensions) is expected to explain such a delay. The integral equation (\ref{eq:integ}), on the other hand, naturally provides a delay. 

The simplest possibility is to assume that $R_0$ decreases instantaneously after a successful lockdown. The left panel of Figure \ref{fig:evolution} shows how the number of reported new cases evolves after a lockdown happens on March 13 (day 73 of 2020) which reduces $R_0$ to 0.95. A more realistic situation is shown in the right panel of Figure \ref{fig:evolution}. Here the value of $R_0$ was gradually decreased from March 11 to March 14. 
In both cases the expected delay, which is clearly visible in the data, is only explained by the integral equation. The reported data of many countries show an interesting oscillation after the effect of a lockdown starts to show up. This feature is also naturally described by the integral equation. 

One remark is in order. The incubation period of 6.4 days is accidentally very close to the weakly cycle of 7 days. Thus, the oscillation might in principle be just a weekend effect. Looking at the weekend during the strongly exponential growth with enough statistics (March 14, Saturday to March 16, Monday) one observes less cases during Saturday and Sunday and an accordingly higher number of cases on Monday than the average exponential growth would predict.
Correcting the data for this effect later would weaken but not eliminate the observed oscillation, it seems to be a real effect. This is also supported by the fact that different countries have minima and maxima of their oscillation on different days (e.g. in Italy the minima are on Mondays and Tuesdays, in the Netherlands they are on Tuesdays and Wednesdays).

It is in principle possible to determine the incubation time $t_e$ and $t_i$ from this oscillatory pattern.
Taking the distance of subsequent minima we determined $t_e+t_i$. Using data of New York City, Italy, Spain, Germany, and the Netherlands~\cite{nyc_data,ecdc} the period of the oscillations seems quite robust and is around $t_e+t_i=7.4$ days on average with a spread of 0.2 days. 

\begin{figure}
\begin{center}
\includegraphics[width=0.7\textwidth]{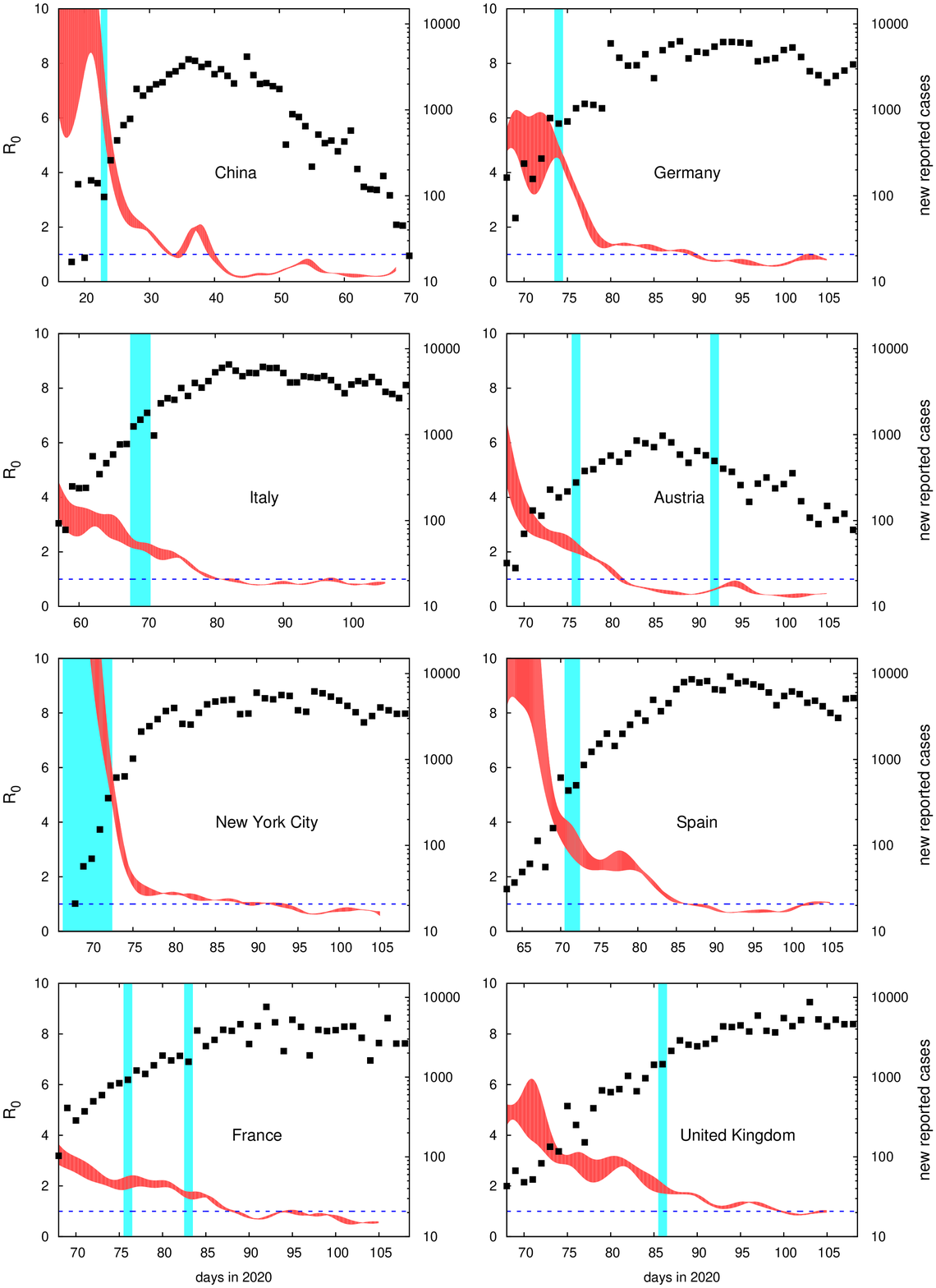}
\caption{
Determination of the basic reproduction number $R_0$ as a function of time for China, Germany, Italy, Austria, New York City, Spain, France and the United Kingdom. We used the integral equation (\ref{eq:integ}) to compute $R_0$ from the available data~\cite{ecdc,austria_data,nyc_data}. The red bands show the results. We varied $t_i$ between 1 and 2 days and $t_e+t_i$ between 6.5 and 7.5 days. The uncertainty associated with this variation is shown by the widths of the bands. As an illustration the cyan areas show the approximate time periods when well defined restrictive measures were introduced within a short time interval (lockdown of Hubei province in China (January 23), lockdown of Lombardy (March 8) and the whole of Italy (March 10), closing schools in Germany (March 14), 
nationwide curfew (March 16) and compulsory wearing of masks (April 1) in Austria, state of emergency (March 7) and closing of Broadway theatres (March 12) in New York City, closing schools in Spain (March 11-12), closing schools (March 16) and restriction of movement (March 23) in France, and issuing statutory instrument 350 in the UK (March 26)). For completeness the original data are also shown (right vertical axes).
Clearly as health authorities are getting better and better data the accuracy on $R_0$ determination will improve.
\label{fig:r0_countries}
}
\end{center}
\end{figure}

Once $t_e$ and $t_i$ are known, one can solve eqn~(\ref{eq:integ}) for $\beta(t)$ or equivalently $R_0(t)$ by taking $\rho(t)$ and the (numerical) integral from the actual data:
\begin{equation}
R_0(t)=\frac{t_i\rho(t)}{\int_{t-t_e-t_i}^{t-t_e}\rho(\tau)d\tau}.
\end{equation}
In this way one can continuously monitor the effect of restrictive or easing measures.  We used this procedure to analyse data from seven countries as well as New York City. In each country the raw data for the daily number of new infections shows large daily fluctuations, most likely due to some anomalies in reporting
new cases.
In order to remove these fluctuations we performed the following smearing procedure. If the original data is given by $\bar{\rho}(t)$ then
we replace this by $w\bar{\rho}(t-1)+(1-2w)\bar{\rho}(t)+w\bar{\rho}(t+1)$ and repeat this procedure three times. We checked that the function $R_0(t)$ depends only mildly on the choice of $w$. The results, for which we used $w=0.25$ are shown in Figure~\ref{fig:r0_countries}.

For all regions we can observe a large drop in $R_0$. However, while in China $R_0$ went significantly below 1, effectively stopping the epidemic, in European countries and New York City it is just below or around 1 even today. It seems inevitable that in these regions any significant easing of the restrictions will induce a second wave of the epidemic. An accurate monitoring of $R_0$ is extremely important in this situation. As demonstrated above, the integral equation formulation is the right approach here.

\section*{Conclusions and Outlook}

In the previous sections we compared the SEIR model and the integral equation
when both $t_e$ and $t_i$ are fixed. In reality the incubation period is described by a probability
function $P(\tau)$ which gives the probability that a person is infectious a time $\tau$ after exposure. The integral equation (\ref{eq:integ}) can easily be generalised to include $P$:
\begin{equation}
\rho(t)=\beta\int_{0}^{\infty} \rho(t-\tau)P(\tau) d\tau,\qquad\qquad R_0=\beta\int_0^{\infty}P(\tau)d\tau.
\end{equation} 
In general, the transmission rate $\beta$ also changes in time since new
restrictive measures can be implemented. This can also be incorporated into
the equation, however, in that case $R_0$ might not have a transparent
interpretation.   
This equation has been around in the literature for a long time (see e.g. \cite{hethcote1980integral,
wearing2005appropriate,wallinga2007generation}) but unfortunately it has not yet been widely adopted. Any evolution computed using this equation is expected to have qualitatively similar features (delay and oscillation) as in the simple approximation presented above. The SEIR model is not a good approximation of this integral equation even if the $P(\tau)$ probability is built into its parameters.

For later stages of a pandemic when $S/N<1$ the integral equation can be generalised to include $S$ as well. 
Furthermore, one can divide the population into sub-compartments e.g. by age groups or location. Each sub-compartment with a population of $N_i$ can be described by its own $\rho_i(t)$ infection history and $P_i(\tau)$ probability function and the cross infection rates between compartments $j$ and $i$ by $\beta_{i\leftarrow j}$. The generalised coupled integral equations in this case read:
\begin{equation}
\rho_i(t)=\frac{S_i(t)}{N_i}\sum_j\beta_{i\leftarrow j}\int_{0}^{\infty} \rho_j(t-\tau)P_j(\tau) d\tau,\qquad S_i(t)=N_i-\int_{-\infty}^t \rho_i(\tau)d\tau.
\end{equation}

As an example, we ran some simulations of the full course of the epidemic
until $S$ saturated close to zero. We found that depending on the parameters
the integral equation and the SEIR equations can predict a significantly different
maximal number of cases at the peak of the epidemic. Any further extension which can be
included in the SEIR equations (e.g. birth and death rate, day/night
differences, inhomogeneities, meta-population systems, etc.) can also be
naturally included in the integral equation formalism.

We demonstrated that the approximation leading to differential equation based
models of epidemic spread is not sufficient in the case of the COVID-19
pandemic. In particular, differential equations provide a good description
only if $R_0$ is close to 1 and if it changes only slowly compared to the
incubation period of the disease. Note that even if $R_0$ is larger than 1,
the initial exponential phase of the epidemic can be well described by
differential equations, but they can significantly underestimate the value of
$R_0$, the larger the $R_0$, the more so. If, however, as a result of
restrictive measures, $R_0$ suddenly drops, differential equations fail to
describe the ensuing oscillatory behaviour of the number of new cases and as a
result, they are not suitable for a precise monitoring of $R_0$. In contrast,
integral equations are naturally suited to this task, even if sharp changes
occur in $R_0$. We also presented a simple way to implement the integral
equation formalism in numerical simulations, which do not become more
complicated than the numerical solution of the SEIR equations. In summary, the
integral equation approach has more predictive power than the most widely used
differential equation based models and also eliminates a serious uncontrolled
approximation of the latter. In the light of the present results we urge
practitioners of the field to rethink and possibly consider implementing the integral 
equation based technique.

\bibliography{refs}{}

\end{document}